# MammoFL: Mammographic Breast Density Estimation using Federated Learning

Ramya Muthukrishnan[1,3], Angelina Heyler[2,4], Keshava Katti[2], Sarthak Pati[5,6,7,8], Walter Mankowski[7,9], Aprupa Alahari[1], Michael Sanborn[2], Emily F. Conant[9], Christopher Scott[10], Stacey Winham[10], Celine Vachon[10], Pratik Chaudhari[2], Despina Kontos[7,9]*, Spyridon Bakas[5,6,7,8]*

**Affiliations:**

[1]Department of Computer and Information Science, University of Pennsylvania, Philadelphia, PA, USA.

[2]Department of Electrical and Systems Engineering, University of Pennsylvania, Philadelphia, PA, USA.

[3]Computer Science and Artificial Intelligence Laboratory (CSAIL), Massachusetts Institute of Technology, Cambridge, MA, USA.

[4]Research and Development, GSK, Cambridge, MA, USA.

[5]Division of Computational Pathology, Department of Pathology and Laboratory Medicine, Indiana University School of Medicine, Indiana, IN, USA.

[6]Research Center for Federated Learning in Medicine, Indiana University School of Medicine, Indiana, IN, USA.

[7]Center for Biomedical Image Computing and Analytics (CBICA), University of Pennsylvania, Philadelphia, PA, USA.

[8]Department of Pathology & Laboratory Medicine, Perelman School of Medicine, University of Pennsylvania, Philadelphia, PA, USA.

[9]Department of Radiology, Perelman School of Medicine, University of Pennsylvania, Philadelphia, PA, USA.

[10]Department of Health Sciences Research, Mayo Clinic, Rochester, MN, USA.

\* Corresponding authors: dkontos@upenn.edu, spbakas@iu.edu

# Abstract

In this study, we automate quantitative mammographic breast density estimation with neural networks and show that this tool is a strong use case for federated learning on multi-institutional datasets. Our dataset included bilateral CC-view and MLO-view mammographic images from two separate institutions. Two U-Nets were separately trained on algorithm-generated labels to perform segmentation of the breast and dense tissue from these images and subsequently calculate breast percent density (PD). The networks were trained with federated learning and compared to three non-federated baselines, one trained on each single-institution dataset and one trained on the aggregated multi-institution dataset. We demonstrate that training on multi-institution datasets is critical to algorithm generalizability. We further show that federated learning on multi-institutional datasets improves model generalization to unseen data at nearly the same level as centralized training on multi-institutional datasets, indicating that federated learning can be applied to our method to improve algorithm generalizability while maintaining patient privacy.

# Introduction

Breast cancer remains the most frequent cancer among women, with about 1 in 8 women in the United States developing it over the course of their lifetimes [1]. Mammography is highly effective in identifying breast cancers before they become fatal [2]. However, it suffers from relatively poor sensitivities ranging from 75% to 85%, with lowest sensitivity in detecting cancers in women with the densest breast tissue [3]. Mammography screenings consist of two-view (mediolateral oblique (MLO) and craniocaudal (CC) bilateral examinations captured as full-field digital mammography (FFDM) images. Using FFDM images, radiologists visually grade breast density based on the American College of Radiology's Breast Imaging Reporting and Data Systems (BI-RADS). Some radiologists employ simple computer-aided detection (CAD) systems, which generally present limited improvements [4], especially in comparison to their AI-based counterparts [4-7].

Breast density not only limits the sensitivity of mammographic screenings but is also a major risk factor for breast cancer [8]. The most frequent method to grade breast density is using the BI-RADS classification of breast density based on mammographic images [9,10]. This method classifies the breast into one of four density categories defined to grade the degree of potential "masking" of cancers by dense tissue rather than quantifying the percent area or volume of glandular tissue, which would help a radiologist better monitor changes in a patient's breast density that point to heightened breast cancer risk.

Current efforts to automate quantitative breast density estimation from mammographic images come in the form of commercially available software and research tools. The semi-automated thresholding tool Cumulus remains the current gold-standard area-based breast percent density (PD) estimation method. Commercial software for volumetric breast composition measurement, like Quantra [11] and Volpara [12], show strong association with Cumulus [13], but they estimate breast PD from x-ray beam interaction models, which make underlying assumptions on metadata that can lead to inaccurate results [7], especially if some of the metadata are not available. Such commercial software may be expensive and suffer from limited interpretability, as the tools do not output a spatial map delineating the dense tissue from the non-dense tissue in the mammogram. Research tools [14-24] come with their own set of limitations. First, except for a few examples such as ImageJ and LIBRA [14], these tools are not freely available, which limits their utility as well as the ability to benchmark their performance. Second, these tools have been trained on small, single-institution datasets [7], which may limit model generalization.

Convolutional neural networks (CNN) have become the workhorse for fully automated medical image analysis tools [25-27]. Neural networks rely on sufficiently large and diverse datasets for training, which are difficult to obtain in the medical field. Single-institution datasets are not sufficient to provide a representative sample for model training and can lead to poor generalization [28], while centrally shared patient data from multiple institutions present privacy and ownership concerns. One solution is federated learning (FL) [28, 29], in which each epoch of network training is distributed to each data owner and then aggregated into a single network, allowing training to leverage data across multiple institutions without data sharing.

We present MammoFL, a tool that federates training of deep neural networks for quantitative PD estimation. MammoFL utilizes two U-Nets to separately segment the breast and dense tissue from the mammogram to estimate PD. In contrast to Deep-LIBRA [3], which segments the dense tissue using traditional methods, MammoFL is uniquely an end-to-end CNN pipeline for PD estimation. We trained MammoFL with FL using the Open Federated Learning (OpenFL) library [30], on labels generated by the publicly available LIBRA [14] tool. We demonstrate that quantitative breast PD estimation is a strong contender for FL on multi-institutional datasets.

| Dataset Type | Training and Validation | | Test | |
|---|---|---|---|---|
| Institution | MC | UPHS | MC | UPHS |
| **Number of Images** | 6,713 | 1,147 | 6,464 | 278 |
| **Number of Women** | 1,679 | 575 | 1,628 | 110 |
| **Screening Start Date** | 2008 | 2003 | 2008 | 2011 |
| **Screening End Date** | 2011 | 2006 | 2014 | 2014 |
| **White (%)** | 98 | 47 | 94 | 21 |
| **Black/Other (%)** | 2 | 53 | 6 | 79 |

**Table 1. General characteristics corresponding to the datasets.** Each dataset is accompanied by its institution, the number of FFDM images present, the number of women in the screening cohort, the year that the screening began, the year that the screening finished, the percent of white subjects present in the screening and the percent of black/other subjects present in the screening.

## Materials and Methods

### Study Datasets

The dataset used for network training and validation (**Table 1**) consisted of non-actionable FFDM screening exams obtained from the University of Pennsylvania Health System (UPHS), Philadelphia, PA, and the Mayo Clinic (MC), Rochester, MN. It included 6,713 bilateral CC-view images from 1,679 women from the MC and 1,147 bilateral MLO-view images from 575 women from UPHS. Each subject was randomly sorted into non-overlapping training and validation sets with an 80:20 split. The ground-truth breast tissue and dense tissue segmentations in this study were obtained by running the LIBRA algorithm [14] on these mammograms. LIBRA first segments the breast from the background in the image and then uses the image of the segmented breast as an input for a second segmentation task to locate the dense versus the fatty tissue. PD labels were calculated by dividing the dense tissue area by the breast tissue area. The holdout test dataset consisted of similar data from UPHS and the MC: 278 MLO-view images from 110 women at UPHS and 6,463 bi-lateral CC and MLO images from 1,628 women at the MC.

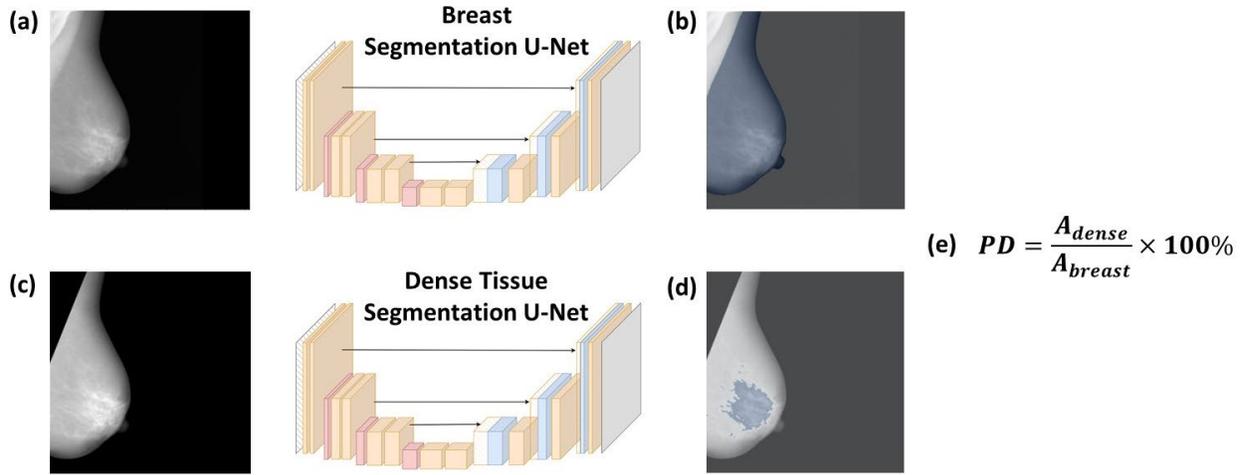

**Figure 1. Network architecture. (a)** This image shows a pre-processed mammogram in MLO-view from the UPHS dataset. Image pre-processing includes removal of the metal tag, pixel normalization, resizing, and contrast adjustment. **(b)** The breast segmentation mask predicted by the first U-Net is displayed for this example. This U-Net segments the breast from the pre-processed image, including the background and pectoralis muscle. This segmentation is used to estimate the breast area. **(c)** The breast mask from (b) is used to remove the background and pectoralis muscle from the image entirely, and the resulting image is the input to the second U-Net. During training, the ground-truth breast mask is used; during inference, the predicted breast mask is used. **(d)** The dense tissue segmentation mask predicted by the second U-Net is displayed for this example. This U-Net segments the dense tissue from the non-dense tissue, and the resulting segmentation is used to estimate the dense tissue area. **(e)** The estimated dense tissue area is divided by the estimated breast area, representing the proportion of dense tissue detected in the breast. This value is the estimated PD value of the CNN pipeline.

**Data Pre-Processing**

Each original mammogram, stored as a DICOM image, was pre-processed by removing the metal tag in the image, down-sampling the images to ensure a standardized shape across the cohort (i.e., 512x512), rescaling pixel intensities to [0,1] using min-max scaling, and removing the metal tag in the mammogram. These images are the inputs to the breast segmentation model during training and inference (**Figure 1a**). The inputs to the dense tissue segmentation model are further pre-processed by removing non-breast pixels (determined by the ground-truth breast mask during training and the predicted breast mask during inference) and again re-normalizing pixel values with min-max scaling (**Figure 1c**).

**Model Architecture**

MammoFL consists of two U-Nets [31], the first for identifying the breast from the entire mammogram (**Figure 1b**) and the second for delineating the dense tissue region from the breast (**Figure 1d**). The two networks have identical architectures but different weights, as they are separately trained on different tasks. The U-Net encoder uses the ResNet34 backbone [32]. During evaluation and inference, predicted segmentations from each U-Net are resampled to the original image size. The number of pixels in each predicted segmentation are calculated as a measure of area, with the area of the dense tissue region divided by the area of the breast region returning the breast PD (**Figure 1e**).

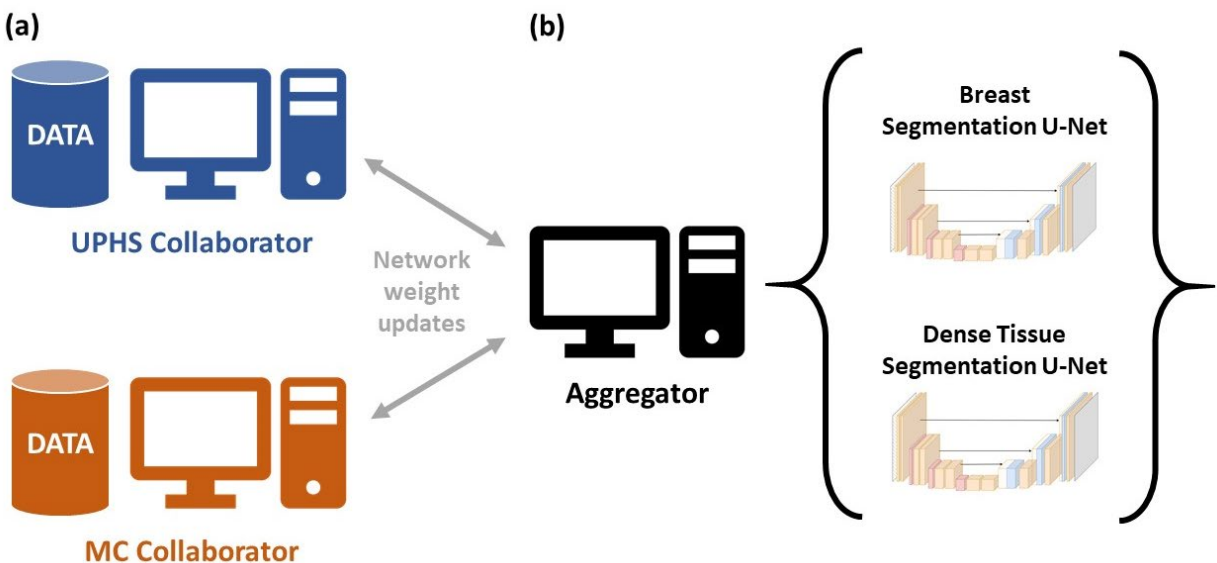

**Figure 2. Workflow for federated training on multi-institution data. (a)** Our FL simulation involves data from two different institutions, UPHS and MC. Each institution's data is housed on a collaborator machine, which locally trains the network and then transfers network updates to the aggregator machine via a secure connection, preventing cross-institutional data sharing. **(b)** The aggregator maintains the final trained U-Net architectures for breast and dense tissue segmentation, respectively, by taking a weighted average of network updates from each collaborator.

**Federated Learning**

We employ the aggregator-server FL framework [33] using OpenFL. Since UPHS and MC are separate institutions or data owners, during each epoch of training, the network weights are updated by training separately on each institution's data, and then aggregated into one set of weights. The aggregation is a weighted average of each institution's updated network, in which the averaging weights are directly proportional to each single-institution dataset size. This simulates a scenario in which each institution's data is securely housed in institution-owned "collaborator" machines, and an "aggregator" machine aggregates each institution's network updates into a single trained network. As a result, only network weights are shared between machines during network training, avoiding cross-institutional data sharing. This simulated FL scenario is visualized in **Figure 2**.

**Model Training**

Both U-Nets were trained with the same hyperparameters, with a batch size of 16, a learning rate of 1e-4, and weight decay of 1e-4. The models were trained for 30 epochs with the Adam optimizer. Data augmentation (random spatial transformations, such as flipping) was used to improve model generalization. The dataset was randomly split into training and validation datasets using a 4:1 ratio. Validation performance was used for hyperparameter optimization.

**Model Evaluation**

PD estimation was evaluated by calculating the mean absolute error (MAE) and Spearman's correlation coefficient between the ground-truth and estimated PD values. However, neither metric spatially captures segmentation accuracies. Thus, the Dice-Sorensen coefficient (DSC) [34], a widely used performance metric for segmentation tasks [35], was used to evaluate the performance of each individual segmentation network with respect to the LIBRA-generated ground-truth labels. DSC measures the area of overlap between ground-truth segmentations $X$ and algorithmic segmentations $Y$. It is computed by the equation $DSC = \frac{2|X \cap Y|}{|X|+|Y|}$. The DSC value ranges from 0 to 1, with 1 denoting perfect overlap of the ground truth and predicted segmentations. Reported DSC values are averaged across images.

To analyze the performance of our algorithm trained with FL, we trained three non-federated ("centralized" training) baselines on: (1) only UPHS data, (2) only MC data, and (3) both UPHS and MC data. We used a paired statistical test to compare all performance metrics between MammoFL and all three baselines by subject. Because some distributions of performance differences were not approximately normal, the Wilcoxon signed-rank test was used, as it is a nonparametric paired test. Since the validation data was used for hyperparameter optimization, the algorithm is evaluated on the holdout test dataset to provide a fair assessment of algorithm performance.

We additionally evaluated how well each model's PD predictions correlate with gold-standard manually-derived Cumulus PDs. We only had access to these Cumulus labels for CC view images in the MC training/validation datasets. We report PD correlation values on the MC validation dataset, using Spearman's correlation coefficient.

**Code Availability**

All code related to network design, breast PD estimation, and federated training can be found at: https://github.com/ramyamut/MammoFL.

| Training Type | Training/ Validation Data | Performance on MC Test Data | | | Performance on UPHS Test Data | | |
|---|---|---|---|---|---|---|---|
| | | MAE (%) | Breast DSC | Dense DSC | MAE (%) | Breast DSC | Dense DSC |
| Centralized | MC | 4.1659 ± 5.4914 | 0.9784 ± 0.0868 | 0.7503 ± 0.1784 | 16.5408 ± 17.4046 | 0.1418 ± 0.1228 | 0.0071 ± 0.0669 |
| Centralized | UPHS | 5.2044 ± 4.9965 | 0.9522 ± 0.0352 | 0.6638 ± 0.208 | 4.1695 ± 4.0902 | 0.9724 ± 0.0302 | 0.6431 ± 0.2054 |
| Centralized | UPHS + MC | 3.7206 ± 4.2591 | 0.9902 ± 0.0203 | 0.7665 ± 0.1465 | 3.4971 ± 4.0069 | 0.9722 ± 0.0287 | 0.6846 ± 0.2021 |
| **Federated** | **UPHS + MC** | **4.2402 ± 4.3948** | **0.9881 ± 0.0229** | **0.7637 ± 0.1656** | **3.9586 ± 3.8566** | **0.9657 ± 0.0304** | **0.6417 ± 0.2268** |

**Table 2. Performance on holdout test dataset using federated versus centralized training.** As expected, the models trained on both HUP and MC data perform better on both institutions' data than those trained on single-institution datasets, motivating the need for multi-institutional collaboration for this problem. MammoFL's performance closely approaches that of the baseline trained on both institutions' data while also preventing cross-institutional data sharing.

| Baseline Training Type | Baseline Training Data | p-value of Federated Model Performance vs. Baseline on MC Test Data | | | p-value of Federated Model Performance vs. Baseline on UPHS Test Data | | |
|---|---|---|---|---|---|---|---|
| | | MAE | Breast DSC | Dense DSC | MAE | Breast DSC | Dense DSC |
| Centralized | MC | 0.9060 | <0.0001 | <0.0001 | <0.0001 | <0.0001 | <0.0001 |
| Centralized | UPHS | <0.0001 | <0.0001 | <0.0001 | 0.1680 | 1.000 | 0.5452 |
| Centralized | UPHS + MC | <0.0001 | <0.0001 | 0.0271 | 0.0110 | <0.0001 | 0.0001 |

**Table 3. p-values of performance differences between MammoFL and baselines on the holdout test dataset.** When evaluated on MC data, MammoFL outperforms the baseline trained on only UPHS data, and when evaluated on UPHS data, it outperforms the baseline trained on only MC data ($p < 0.05$). However, the baseline trained on both institutions' data outperforms the federated model on all test dataset evaluation metrics ($p < 0.05$), demonstrating current limitations of FL and the tradeoff between generalizability and patient privacy.

## Results

**Evaluation on Independent Test Dataset**

The model trained on both datasets with centralized learning resulted in PD MAEs of 3.7206 ± 4.2591 and 3.4971 ± 4.0069, PD correlations of 0.7977 and 0.7893, breast segmentation DSCs of 0.9902 ± 0.0203 and 0.9722 ± 0.0287, and dense tissue segmentation DSCs of 0.7665 ± 0.1465 and 0.6846 ± 0.2021, on the MC and UPHS holdout test datasets, respectively. MammoFL resulted in PD MAEs of 4.2402 ± 4.3948 and 3.9586 ± 3.8566, PD correlations of 0.7833 and 0.7398, breast segmentation DSCs of 0.9881 ± 0.0229 and 0.9657 ± 0.0304, and dense tissue segmentation DSCs of 0.7637 ± 0.1656 and 0.6417 ± 0.2268, on the MC and UPHS test datasets, respectively. As expected, the models trained on single-institution datasets generalized poorly to the other institution's test data, compared to the models that utilized both datasets for training and validation. Furthermore, the PD predictions of both models trained on multi-institution data consistently showed greater correlation with LIBRA PD labels than either single-institution baseline. Interestingly, for several metrics, such as PD correlation and segmentation DSCs for the MC data, both models trained on multi-institution data outperformed the single-institution baselines on test data from the same institution, suggesting that training on multi-institution datasets is not only critical to generalization to other institutions' data but also unseen data from the same institution. Correlation plots are displayed in **Figures 3 and 4**, and all other results are listed in **Table 2**.

MammoFL outperforms the single-institution models on all evaluation metrics ($p < 0.05$) when evaluated on the other institution's test data (**Table 3**). However, there is also a statistically significant difference in all evaluation metrics between MammoFL and the centralized baseline trained on both institutions' data, with the latter outperforming the former. This result highlights the limitations of FL compared to standard, centralized training on multi-institution data.

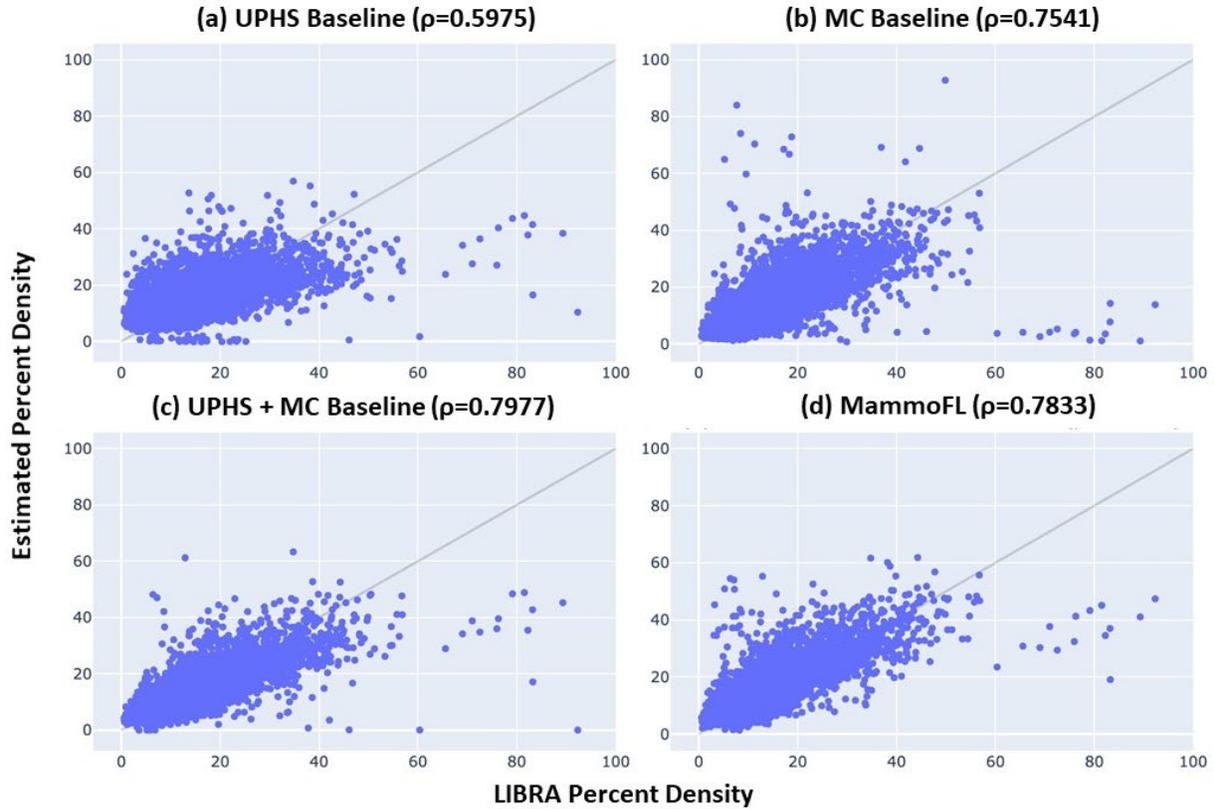

**Figure 3. LIBRA PDs vs. MammoFL/baseline PDs on MC test data.** When evaluated on unseen MC data, both models trained on multi-institution data show higher correlation with PD labels than the models trained on single-institution data, including the model trained only on MC data. This result exemplifies the importance of training on multi-institution data to generalization to unseen data and demonstrates that MammoFL's performance is close to that of the baseline trained on multi-institution data. All correlation values are statistically significant ($p < 0.05$).

**Evaluation on Gold-Standard Cumulus Labels**

The LIBRA-generated PD labels are moderately correlated with the Cumulus PDs with a correlation of $\rho=0.7010$. The correlations between the non-federated baselines' estimated PD values and the Cumulus PD labels are: $\rho=0.8104$ for the baseline trained on both UPHS and MC data, $\rho=0.5926$ for the baseline trained only on UPHS data, and $\rho=0.8002$ for the baseline trained only on MC data. The latter result is expected since it is not evaluating generalizability; networks trained on a single-institution dataset are expected to perform well on data from the same institution. The correlation metric for MammoFL was $\rho=0.7703$, demonstrating the ability of federated training to closely approach the performance of centralized training on multi-institution datasets. All results are statistically significant (p < 0.05) and displayed in **Table 4**.

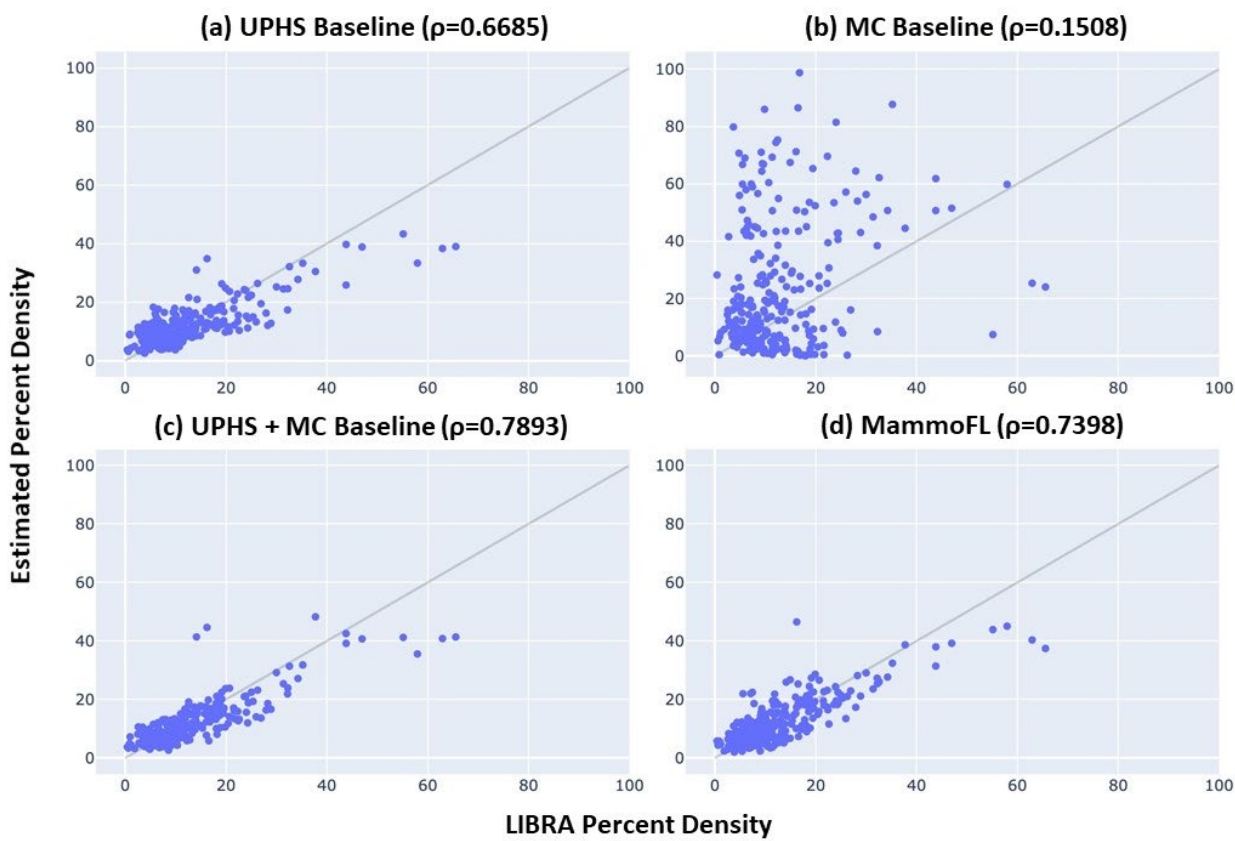

**Figure 4. LIBRA PDs vs. MammoFL/baseline on UPHS test data.** The trends demonstrated in Figure 3 are replicated here, further making the case that training on multi-institution datasets is necessary for strong generalization capabilities, and that FL achieves similar performance to that of centralized training on multi-institution data. All correlation values are statistically significant (p < 0.05).

| Training Type | Training Data | Spearman correlation coefficient with Cumulus PDs ($\rho$) | p-value of $\rho$ | 95% confidence interval of $\rho$ |
|---|---|---|---|---|
| Synthetic LIBRA labels | | 0.7012 | <0.0001 | [0.6334, 0.7573] |
| Centralized | MC | 0.8002 | <0.0001 | [0.7709, 0.8263] |
| Centralized | UPHS | 0.5926 | <0.0001 | [0.5404, 0.6402] |
| Centralized | UPHS + MC | 0.8104 | <0.0001 | [0.7824, 0.8352] |
| Federated | UPHS + MC | 0.7703 | <0.0001 | [0.7371, 0.7997] |

**Table 4. Correlations of algorithm-estimated PDs and Cumulus gold-standard PDs in MC validation dataset.** The Spearman correlation coefficient between gold-standard Cumulus PD labels and model predictions of PD is calculated for MammoFL and each centralized baseline, as well as the synthetic LIBRA labels. Among these, all models trained on MC data show the greatest correlations with Cumulus PDs.

**Discussion**

In this study, we train and evaluate an end-to-end CNN pipeline to estimate breast PD, a quantitative measure of breast density. Furthermore, we show that training on multi-institutional datasets is important for better algorithm generalizability for this problem, and that FL can be used to successfully achieve this. This allows our algorithm to reap the benefits of training on larger, multi-institutional datasets while maintaining patient privacy. Our work demonstrates proof-of-concept that quantitative breast PD estimation with CNNs is a strong use case for FL in medicine.

The results demonstrate the need to train on multi-institutional datasets. We first show the problem that although models trained on solely UPHS or MC data performed well on test data from the same source, these models did not generalize well to data from a different institution. That is, the UPHS model generalized poorly to MC data, and vice versa. We next demonstrate that models trained on multi-institution data can outperform single-institution models on test data from the same source and generalize to other institutions' data significantly better. Both MammoFL and the multi-institution baseline show greater correlation with ground-truth PD values than either single-institution baseline, when evaluated on test data from the same institution. Both also outperform the baseline trained only on MC data on breast and dense tissue DSC, when evaluated on MC test data. With regard to generalization performance, MammoFL outperformed both single-institution baselines across almost all metrics measured in the test dataset. This demonstrates that training models on multi-institution datasets enables better model generalizability.

The results also show that the model trained with FL reaches comparable performance as the model trained with a standard centralized training scheme, since for most metrics, MammoFL's performance lies between that of the multi-institution baseline and those of the single-institution baselines. However, the multi-institution baseline consistently outperformed MammoFL on all metrics with statistical significance. Thus, there is still a performance gap between the two methods, emphasizing the tradeoffs between privacy and generalization performance introduced by FL while also leaving room for future work to design more robust federated training algorithms.

One limitation of this work is that the ground-truth labels used for algorithm development were synthetically generated by the LIBRA algorithm rather than gold-standard manual labels. This choice was made because Cumulus gold-standard PD labels were not available for the UPHS dataset, which is required to train and evaluate algorithms on a multi-institutional dataset. Since the LIBRA algorithm has been shown to accurately predict gold-standard PD values, we hope that this work is a proof-of-concept demonstration of the utility of FL for mammographic PD estimation. To partly address this limitation, we analyze the correlations between trained models' predictions and the Cumulus PD labels of the MC validation data's CC view images. The multi-institution baseline showed the greatest correlation with these gold-standard labels, while the federated model's correlation was comparable. We see that the single-institution baselines' performance trends are replicated here: the model trained on MC data performs well on the MC validation data, while the UPHS data generalizes poorly. With the exception of the UPHS model, the results also demonstrate that all CNN-based architectures outperform LIBRA, as these models show a greater correlation with the Cumulus PDs than the LIBRA labels used for training. This may be due to CNNs' abilities to robustly identify relevant patterns to the task at hand despite noisy labels. Future work will incorporate Cumulus gold-standard labels for FL, as these labels are more accurate than the LIBRA-generated labels and thus provide for a more grounded analysis of the utility of FL for breast cancer screenings.

Additionally, women with large breasts require multiple overlapping CCs and MLOs for complete breast imaging. However, the cases in this study were only chosen from women who had standard views. We leave it to future work to handle multiple views and account for overlapping areas of dense tissue between views.

We created MammoFL, a neural network pipeline to calculate breast PD from mammograms. We demonstrate that training on multi-institutional datasets is necessary for generalizability, and that FL allows the algorithm to preserve privacy while maintaining the benefits of training on diverse datasets. We have made our code publicly available in hopes that our tool will allow researchers to train our networks on large, multi-institutional datasets, accelerating breast cancer research and clinical care.

# Acknowledgments

This work was supported in part by the University of Pennsylvania Department of Electrical and Systems Engineering Senior Design Fund.

# References

1. Howlader, N., Noone, A.M., Krapcho, M., Miller, D., Brest, A., Yu, M., Ruhl, J., Tatalovich, Z., Mariotto, A., Lewis, D.R., Chen, H.S., Feuer, E.J., Cronin, K.A. SEER Cancer Statistics Review, 1975-2016, *National Cancer Institute* (2019).

2. Duffy, S.W., Tabár, L., Yen, A.M.-F., Dean, P.B., Smith, R.A., Jonsson, H., Törnberg, S., Chen, S.L.-S., Chiu, S.Y.-H., Fann, J.C.-Y., Ku, M.M.-S., Wu, W.Y.-Y., Hsu, C.-Y., Chen, Y.-C., Svane, G., Azavedo, E., Grundström, H., Sundén, P., Leifland, K., Frodis, E., Ramos, J., Epstein, B., Åkerlund, A., Sundbom, A., Bordás, P., Wallin, H., Starck, L., Björkgren, A., Carlson, S., Fredriksson, I., Ahlgren, J., Öhman, D., Holmberg, L., Chen, T.H.-H. Mammography screening reduces rates of advanced and fatal breast cancers: Results in 549,091 women. *Cancer* **126**, 2971-2979 (2020).

3. Domingo, L., Hofvind, S., Hubbard, R.A., Roman, M., Benkeser, D., Sala, M., Castells, X. Cross-national comparison of screening mammography accuracy measures in U.S., Norway, and Spain. *European Radiology* **26**, 2520-2528 (2016).

4. Maghsoudi, O.H., Gastounioti, A., Scott, C., Pantalone, L., Wu, F., Cohen, E.A., Winham, S., Conant, E.F., Vachon, C., Kontos, D. Deep-LIBRA: An artificial-intelligence method for robust quantification of breast density with independent validation in breast cancer risk assessment. *Medical Image Analysis* **73**, 102138 (2021). Sprague, B.L., Gangnon, R.E., Burt, V., Trentham-Deitz, A., Hampton, J.M., Wellman, R.D., Kerlikowske, K., Miglioretti, D.L. Prevalence of mammographically dense breasts in the United States. *Journal of the National Cancer Institute* **106** (10) (2014).

5. Mohamed, A.A., Berg, W.A., Peng, H., Luo, Y., Jankowitz, R.C., Wu, S. A deep learning method for classifying mammographic breast density categories. *Medical Physics* **45** (1), 314-321 (2018).

6. Lehman, C.D., Yala, A., Schuster, T., Dontchos, B., Bahl, M., Swanson, K., Barzilay, R. Mammographic Breast Density Assessment Using Deep Learning: Clinical Implementation. *Radiology* **290** (1), 52-58 (2019).

7. Van der Velden, B.H.M., Janse, M.H.A., Ragusi, M.A.A., Loo, C.E., Gilhuijs, K.G.A. Volumetric breast density estimation on MRI using explainable deep learning regression. *Scientific Reports* **10** (18095) (2020).

8. Sprague, B.L., Gagnon, R.E., Burt, V., Trentham-Deitz, A., Hampton, J.M., Wellman, R.D., Kerlikowske, K., Miglioretti, D.L. Prevalence of mammographically dense breasts in the United States. *Journal of the National Cancer Institute* **106** (10) (2014).

9. Sprague, B.L., Conant, E.F., Onega, T., Garcia, M.P., Beaber, E.F., Herschorn, S.D., Lehman, C.D., Tosteson, A.N.A., Lacson, R., Schnall, M.D., Kontos, D., Haas, J.S., Weaver, D.L., Barlow, W.E., PROSPR Consortium. Variation in mammographic breast density assessments among radiologists in clinical practice: a multicenter observational study. *Annals of Internal Medicine* **165** (7), 457-464 (2016).

10. Irshad, A., Leddy, R., Ackerman, S., Cluver, A., Pavic, D., Abid, A., Lewis, M.C. Effects of changes in BI-RADS density assessment guidelines (fourth versus fifth edition) on breast density assessment: intra-and interreader agreements and density distribution. *American Journal of Roentgenology* **207** (6), 1366-1371 (2016).

11. Hartman, K., Highnam, R., Warren, R., Jackson, V. Volumetric Assessment of Breast Tissue composition from FFDM Images. *International Workshop on Digital Mammography* **5116**, 33-39 (2008).

12. Highnam, R., Brady, S.M., Yaffe, M.J., Karssemijer, N., Harvey, J. Robust Breast Composition Measurement. *International Workshop on Digital Mammography* **6136**, 378-385 (2010).

13. Kontos, D. Bakic, P.R., Acciavatti, R.J., Conant, E.F., Maidment, A.D.A. A Comparative Study of Volumetric and Area-Based Breast Density Estimation in Digital Mammography: Results from a Screening Population. *International Workshop on Digital Mammography* **6136**, 378-385 (2010).

14. Keller, B.M., Nathan, D.L., Wang, Y., Zheng, Y., Gee, J.C., Conant, E.F., Kontos, D. Estimation of breast percent density in raw and processed full field digital mammography images via adaptive fuzzy c-means clustering and support vector machine segmentation. *Medical Physics* **38** (8), 4903-4917 (2012).

15. Mustra, M., Grgic, M., Rangayyan, R.M. Review of recent advances in segmentation of the breast boundary and the pectoral muscle in mammograms. *Medical & Biological Engineering & Computing* **54**, 1003-1024 (2016).

16. Li, Y., Chen, H., Yang, Y., Yang, N. Pectoral muscle segmentation in mammograms based on homogenous texture and intensity deviation. *Pattern Recognition* **46** (3), 681-691 (2013).

17. Shi, P., Zhong, J., Rampun, A., Wang, H. A hierarchical pipeline for breast boundary segmentation and calcification detection in mammograms. *Computers in Biology and Medicine* **96**, 178-188 (2018).

18. Ferrari, R.J., Rangayyan, R.M., Desautels, J.E.L., Borges, R.A., Frere, A.F. Automatic identification of the pectoral muscle in mammograms. *IEEE Transactions on Medical Imaging* **23** (2), 232-245 (2004).

19. Mustra, M., Grgic, M. Robust automatic breast and pectoral muscle segmentation from scanned mammograms. *Signal Processing* **93** (10), 2817-2827 (2013).

20. Nagi, J., Kareem, S.A., Nagi, F., Ahmed, S.K. Automated breast profile segmentation for ROI detection using digital mammograms. *2010 IEEE EMBS Conference on Biomedical Engineering and Sciences (IECBES)*. IEEE, 87-92 (2010).

21. Taghanaki, S.A., Liu, Y., Miles, B., Hamarneh, G. Geometry-based pectoral muscle segmentation from MLO mammogram views. *IEEE Transactions on Biomedical Engineering* **64** (11), 2662-2671 (2017).

22. Rampun, A., Morrow, P.J., Scotney, B.W., Winder, J. Fully automated breast boundary and pectoral muscle segmentation in mammograms. *Artificial Intelligence in Medicine* 79, 28-41 (2017).

23. Czaplicka, K., Włodarczyk, J. Automatic breast-line and pectoral muscle segmentation. *Schedae Informaticae* **20** (2011).

24. Dembrower, K., Liu, Y., Azizpour, H., Eklund, M., Smith, K., Lindholm, P., Strand, F. Comparison of a deep learning risk score and standard mammographic density score for breast cancer risk prediction. *Radiology* **294** (2), 265-272 (2020).

25. Anwar, S.M., Majid, M., Qayyum, A., Awais, M., Alnowami, M., Khan, M.K. Medical image analysis using convolutional neural networks: a review. *Journal of Medical Systems* **42** (226), (2018).

26. Sarvamangala, D.R., Kulkarni, R.V. Convolutional neural networks in medical image understanding: a survey. *Evolutionary Intelligence* **15**, 1-22 (2021).

27. Lu, L., Zheng, Y., Carneiro, G., Yang, L. Deep learning and convolutional neural networks for medical image computing. *Advances in Computer Vision and Pattern Recognition* (2017).

28. Sheller, M.J., Edwards, B., Reina, G.A., Martin, J., Pati, S., Kotrotsou, A., Milchenko, M., Xu, W., Marcus, D. Colen, R.R., Bakas, S. Federated learning in medicine: facilitating multi-institutional collaborations without sharing patient data. *Scientific Reports* **10**, 12598 (2020).

29. Pati, S., Baid, U., Edwards, B., Sheller, M., et al. Federated Learning Enables Big Data for Rare Cancer Boundary Detection. *Nature Communications* (2022).

30. Foley, P., Sheller, M.J., Edwards, B., Pati, S., Riviera, W., Sharma, M., Moorthy, P.N., Wang, S., Martin, J., Mirhaji, P., Shah, P., Bakas, S. OpenFL: the open federated learning library. *Physics in Medicine and Biology* (2022).

31. Ronneberger, O., Fischer, P., Brox, T. U-Net: convolutional networks for biomedical image segmentation. *International Conference on Medical Image Computing and Computer-Assisted Intervention*, 234-241 (2015).

32. He, K., Zhang, X., Ren, S., Sun, J. Deep residual learning for image recognition. *Proceedings of the IEEE Conference on Computer Vision and Pattern Recognition (CVPR)*, 770-778 (2016).

33. Reike, N., Hancox, J., Li W., Milletari, F., Roth, H.R., Albarqouni, S., Bakas, S., Galtier, M.N., Landman, B.A., Maier-Hein, K., Ourselin, S., Sheller, M., Summers, R.M., Trask, A., Xu, D., Baust, M., Cardoso, M.J. The future of digital health with federated learning. *npj Digital Medicine* **3** (119), (2020).

34. Zijdenbos, A.P., Dawant, B.M., Margolin, R.A., Palmer, A.C. Morphometric analysis of white matter lesions in MR images: method and validation. *IEEE Transactions on Medical Imaging* **13** (4), 716-725, (1994).

35. Maier-Hein, L., Reinke, A., Godau, P., Tizabi, M.D., et al. Metrics reloaded: Pitfalls and recommendations for image analysis validation. *arXiV* (2022).